# CEST-KAN: Kolmogorov-Arnold Networks for CEST MRI Data Analysis

Jiawen Wang[1#], Pei Cai[1#], Ziyan Wang[1], Huabin Zhang[1], Jianpan Huang[1*]

[1] Department of Diagnostic Radiology, The University of Hong Kong, Hong Kong, China
[#] Authors contributed equally

[*] **Corresponding Author:**

Jianpan Huang, PhD

The University of Hong Kong,

5 Sassoon Road, Pokfulam, Hong Kong, China.

Email: jphuang@hku.hk

Tel: 852-28170373

**Grant Support:** The University of Hong Kong: 109000487, 204610401 and 204610519.

**Submitted to** *Magnetic Resonance in Medicine*

# ABSTRACT

**Purpose:** This study aims to propose and investigate the feasibility of using Kolmogorov-Arnold Network (KAN) for CEST MRI data analysis (CEST-KAN).

**Methods:** CEST MRI data were acquired from twelve healthy volunteers at 3T. Data from ten subjects were used for training, while the remaining two were reserved for testing. The performance of multi-layer perceptron (MLP) and KAN models with the same network settings were evaluated and compared to the conventional multi-pool Lorentzian fitting (MPLF) method in generating water and multiple CEST contrasts, including amide, relayed nuclear Overhauser effect (rNOE), and magnetization transfer (MT).

**Results:** The water and CEST maps generated by both MLP and KAN were visually comparable to the MPLF results. However, the KAN model demonstrated higher accuracy in extrapolating the CEST fitting parameters, as evidenced by the smaller validation loss during training and smaller absolute error during testing. Voxel-wise correlation analysis showed that all four CEST fitting parameters generated by KAN consistently exhibited higher Pearson coefficients than the MLP results, indicating superior performance. Moreover, the KAN models consistently outperformed the MLP models in varying hidden layer numbers despite longer training time.

**Conclusion:** In this study, we demonstrated for the first time the feasibility of utilizing KAN for CEST MRI data analysis, highlighting its superiority over MLP in this task. The findings suggest that CEST-KAN has the potential to be a robust and reliable post-analysis tool for CEST MRI in clinical settings.

# 1 INTRODUCTION

Chemical exchange saturation transfer (CEST) MRI is a molecular imaging technique that enables the detection of low-concentration molecules through the exchange between water protons and exchangeable solute protons.[1-3] It has been extensively utilized for imaging various endogenous contrasts, including amide,[4-7] glutamate,[8-11] creatine,[12-16] and glucose.[17-22] Additionally, the investigation of relayed nuclear Overhauser effects (rNOE) at negative 0-5 ppm from the water signal, apart from the direct CEST effects in the water saturation spectrum (Z-spectrum), has gained significant attention over the past years[23-28]. By leveraging the in vivo biomolecules or compounds with exchangeable protons, CEST MRI holds immense potential for disease diagnosis and treatment monitoring.

Despite its potential for clinical use, there are several limitations that hinder the widespread application of CEST MRI. One of these limitations is the absence of a standardized and universally accepted post-processing method for CEST MRI. Moreover, most of the existing post-processing methods are often complex and require specialized knowledge and expertise. The existing methods include magnetization transfer (MT) ratio asymmetry ($MTR_{asym}$)[5,29], Lorentzian difference analysis (LDA)[24,30], multi-pool Lorentzian fitting (MPLF)[31,32], polynomial and Lorentzian line-shape fitting (PLOF)[33,34], three-point method[27,35] and other methods.[36,37] Among these methods, MPLF is a frequently employed CEST post-processing method. It necessitates a full Z-spectrum acquired by using a continuous-wave (CW) radio frequency (RF) saturation pulse with a low saturation power[32]. This method fits the Z-spectrum by presuming a Lorentzian line shape for each exchanging proton pool. MPLF generally works well, especially at high field strengths where the individual CEST peaks can be well distinguished, but its time-consuming nature poses a challenge to its widespread clinical implementation for 3D human brain CEST dataset.

To accelerate and simplify the CEST post-processing, many deep learning techniques, mostly multi-layer perceptron (MLP)-based models, have been extensively utilized in extracting various CEST fitting parameters from Z-spectra with faster speed than conventional methods in recent years.[38-44] MLP-based models have been recently utilized in CEST magnetic resonance fingerprinting (CEST-MRF), which is a rapid quantitative CEST technique, to speed up the reconstruction process.[45-51] Encouraging results from these studies demonstrated the significant promise of leveraging MLP techniques for CEST analysis. Nevertheless, despite the widespread adoption of MLPs, they exhibit notable drawbacks. In transformer, for instance,

MLPs occupy a substantial portion of the non-embedding parameters and tend to be less interpretable compared to attention layers, at least without the aid of post-analysis tools.

Recently, Kolmogorov-Arnold Networks (KANs)[52] have been proposed as promising alternatives to MLPs. Unlike MLPs with fixed activation functions, KANs replace traditional linear weight matrices with learnable spline functions at each weight parameter. This innovative approach allows KANs to provide smaller yet more accurate models with faster scaling and enhanced interpretability.

In this study, we pioneered the exploration of leveraging KAN for the CEST MRI data analysis (CEST-KAN). We rigorously evaluated the MLP method and the KAN method on the same human CEST dataset and under identical parameter settings, using MPLF analysis as both a demonstration example and a ground truth benchmark. We also compared the performances of MLP and KAN with different layers in predicting multiple CEST fitting parameters. The present study also evaluated the advantages and limitations of the CEST-KAN approach as a potential tool for implementation in clinical settings.

## 2 METHODS

### 2.1 Theory

#### 2.1.1 Multi-pool Lorentzian fitting (MPLF)

In CEST post-processing, Z-spectra are generated by normalizing the CEST source images $M_{sat}(\Delta\omega)$ to the $M_0$ image:

$$Z(\Delta\omega) = \frac{M_{sat}(\Delta\omega)}{M_0}, \tag{1}$$

where $\Delta\omega$ is the frequency offsets of Z-spectra. In MPLF analysis, the CEST peak of each exchange pool (*i*) can be depicted by a Lorentzian line shape:[31]

$$L_i(\Delta\omega) = \frac{A_i}{1 + \left[\frac{\Delta\omega - (\Delta\delta_i + \Delta\delta)}{\Gamma_i/2}\right]^2}, \tag{2}$$

where $A_i$, $\Delta\delta_i$, and $\Gamma_i$ represent the peak amplitude, peak position, and full-width-at-half-maximum of eaxhcange pool *i*, $\Delta\delta$ is the B0 inhomogeneity in ppm. Therefore, the Z-spectra can be fitted by the MPLF method using the following equation:

$$Z_{fit}(\Delta\omega) = 1 - \sum_i^N L_i(\Delta\omega), \tag{3}$$

where N is the total number of exchange pools. Considering the CEST data used in this study were acquired from a clinical 3T MRI scanner where the spectral resolution is relatively low, we used a four-pool Lorentzian fitting model which considers direct water saturation (DS) at 0 ppm, amide at 3.5 ppm, rNOE at –3.5 ppm and MT at –2.5 ppm. MPLF analysis was performed using custom-written code in MATLAB (MathWorks, USA). The initial values and bound conditions for MPLF are given in Table 1.

### 2.1.2 Multi-layer perceptrons

Multi-layer perceptrons (MLPs) refer to fully connected feedforward neural networks which are the foundational blocks of various deep learning models[53-55]. MLPs are based on the universal approximation theorem structuring computations through layered transformations, which can be simplified as:

$$MLP(x) = (W_{L-1} \circ \sigma \circ W_{L-2} \circ \sigma \circ \cdots \circ W_1 \circ \sigma \circ W_0)x, \tag{4}$$

where $x$ denotes the input (Z values in this study), $L$ refers to the total layers of MLP, $W$ and $\sigma$ represent the linear affine transformations and non-linear continuously differentiable activation functions (such as sigmoid or ReLU), respectively. In MLP (Figure 1), each node represents a neuron, and each connection between the nodes represents a layer weight.

### 2.1.3 Kolmogorov-Arnold Network

Kolmogorov-Arnold networks (KANs) are novel fully connected deep neural networks introduced very recently with a radical shift from the MLP paradigm. They are based on Kolmogorov-Arnold theorem, which can be simplified as:[52]

$$KAN(x) = (\phi_{L-1} \circ \phi_{L-2} \cdots \phi_1 \circ \phi_0)x, \tag{5}$$

where x denotes the input (Z values in this study), $L$ refers to the total layers of KAN, and $\phi_l$ is a univariate function corresponding to the $l^{th}$ layer. Unlike MLPs which treat linear transformations and nonlinearities separately as $W$ and $\sigma$, KANs incorporate both together in $\phi$ by simply summing incoming signals transformed by a set of B-spline functions. As a result, there are no linear weight matrices, but learnable activation functions parameterized as splines on edges in each layer of KANs. Intuitively speaking, the summation of these activation functions suggests that the KANs are just combinations of MLPs and splines—featuring an MLP structure externally and a spline design internally. Therefore, this design benefits from the accuracy of splines for low-dimensional functions while suffering less from curse of dimensionality problem owing to its external similarity to MLPs.

## 2.2 MRI scan

The study was approved by the institutional review board of The University of Hong Kong and conducted according to the guidelines. CEST MRI experiments were performed at a 3T MRI system (SIGNA Premier, GE Healthcare) on 12 healthy volunteers with an average age of 28.4 years old (6 males and 6 females, aged 22-41 years old). The CEST sequence consisted of a continuous wave (CW) module with a saturation time of 2 s and a saturation power of 0.8 μT, and a single-shot fast spin echo (FSE) readout module. An $M_0$ image at a frequency offset of −300 ppm and 43 CEST images at frequency offsets ranging from −20 to 20 ppm were acquired. Other parameters were as follows: repetition time (TR) = 3 s, echo time (TE) = 23 ms, field of view (FOV) = 220×220 mm$^2$, acquisition matrix = 128×128, reconstructed matrix size = 282×282, slice thickness = 6 mm. The scan time for each CEST data was 2 minutes 21 seconds.

## 2.3  Network training

For both CEST-MLP and CEST-KAN, the input of networks was the full Z-spectrum of each image voxel in a length of 43, while the output of networks was the 9 CEST fitting parameters derived from the MPLF method, with the pool number set to four, as illustrated in Figure 1. The training and validation of the networks employed CEST data from 10 participants, comprising 260,445 Z-spectra, with 80% (208,356 Z-spectra) used for training and 20% (52,089 Z-spectra) for validation. The remaining CEST data from 2 participants, consisting of 55,368 Z-spectra, was reserved for testing purposes. To ensure a fair comparison, the network size and parameter settings were kept constant. This study involved an investigation and comparison of the CEST-MLP and CEST-KAN networks, with layer numbers ranging from one to four.

During network training, a batch size of 32 was utilized, and the learning rate was set to 0.01 and decayed with a factor of 0.8 every epoch. The optimizer employed was AdamW,[56] a variant of the Adam optimizer with a weight decay (set to $10^{-4}$ in this study). The loss function used was *msereg*, which included a regularization component designed to penalize large weights and biases during training. Specifically, *msereg* was defined as $\gamma \times msw + (1-\gamma) \times mse$, where *msw* represented the mean squared weights of the network, *mse* represented the mean squared error of training, and $\gamma$ was the regularization parameter (set to 0.01 in this study).[44] The early stop patience was set to 50. The network training was conducted using PyTorch on a computer (Intel Core i9, 64 GB memory) with an NVIDIA RTX 4090 graphics processing unit (GPU).

## 3  RESULTS

To compare the training performance of one-layer (100 neurons) MLP and KAN on the same human CEST dataset, the loss curves for both training and validation are displayed in Figure

2. Both networks exhibited fast convergence with loss curves that initially decreased quickly before plateauing. After 50 epochs without improvement in the validation loss, the training process was stopped. The training epochs were almost identical, with MLP and KAN undergoing 104 and 102 epochs, respectively. However, KAN demonstrated obviously lower initial and end losses. The best validation losses for MLP and KAN were 0.463 and 0.241, respectively, indicating that KAN outperformed MLP in predicting the CEST fitting parameters. Although KAN showed promise, it took longer to train than MLP, with training times of 2396 and 1017 seconds, respectively.

To further validate the superiority of KAN over MLP, we applied the trained network models to extract multiple CEST fitting parameters from the unseen testing dataset (2 subjects) and compared them with conventional MPLF. Figure 3 displays the voxel-by-voxel correlation results of two testing subjects between MLP/KAN and MPLF for four (of nine) CEST fitting parameters commonly used to generate water/CEST maps. The higher Pearson coefficients in the four fitting parameters indicated that KAN exhibited better predicting accuracy than MLP (R = 0.998 vs 0.989, 0.979 vs 0.932, 0.987 vs 0.952, and 0.998 vs 0.994 for $A_{water}$, $A_{amide}$, $A_{rNOE}$, and $A_{MT}$, respectively). These quantitative results further supported that KAN had superior performance over MLP in predicting CEST fitting parameters.

CEST MRI offers the advantage of producing CEST maps that contain molecule-related spatial information. We utilized the voxel-basis CEST fitting parameters generated by two network models to reconstruct CEST maps and compared them with MPLF results. Figure 4 displays the water/CEST maps ($A_{water}$, $A_{amide}$, $A_{rNOE}$, and $A_{MT}$) of subject #1 generated by all three methods. Both MLP and KAN produced water/CEST maps that were visually comparable to the results obtained using MPLF (Figure 4A-C). However, KAN demonstrated higher spatial consistency than MLP, as evidenced by smaller errors displayed in the absolute difference maps (Figure 2D&E). Upon closer examination, we observed that the MLP results had a smoothing effect that removed small details in the maps, particularly for $A_{amide}$, $A_{rNOE}$, and $A_{MT}$. In contrast, the KAN results retained these details, providing more spatial information for the CEST maps. These findings were consistently observed in the results of subject #2 (Figure S1).

As the CEST fitting parameters are extrapolated from Z-spectra, it was crucial to compare the performance of the three methods in reversely reconstructing the Z-spectra using their generated CEST fitting parameters. To this end, we selected three regions of interest (ROIs) from different brain regions, including white matter, gray matter, and cerebrospinal fluid, as indicated by green, orange, and blue, respectively in Figure 5A. Figure 5B-C displays the Z-spectra of subject #1 reconstructed using the nine CEST fitting parameters generated by all three methods. The results of MLP/KAN were compared to MPLF, and the absolute difference was calculated. In general, CSF exhibited the sharpest Z-spectra when compared to white matter and gray matter, as fluid-like tissue usually has smaller MT and direct water saturation effects when compared to solid-like tissues. Meanwhile, white matter exhibited a larger MT effect than gray matter, as it is composed of bundles of axons coated with myelin. All three methods captured these apparent features well. However, the comparison revealed that KAN exhibited higher accuracy than MLP in all three regions, as evidenced by the smaller absolute difference when using MPLF as a reference. These observations were also found in the results of subject #2 (Figure S2).

The above presented results proved that one-layer (100 neurons) KAN outperformed the same-size MLP in predicting CEST fitting parameters. To ensure the reliability of these findings, the networks were trained nine more times and the outcomes of the 10 runs for both networks were compared. Additionally, we conducted the comparison experiments of 10 runs for networks with different numbers of layers, ranging from two to four layers with 100 neurons in each layer. The comparison results of validation loss, epoch number, and training time between MLPs and KANs are shown in Figure 6. KANs consistently had lower validation losses than MLPs, and the losses of MLPs were more dependent on the layer number compared to KANs, where the losses were comparable for different layers (Figure 6A). For the epoch number, KAN models converged at a smaller epoch number than MLPs (Figure 6B). However, KANs generally took longer time than MLPs to train, and their training time increased dramatically with the increase in layer number (Figure 6C).

## 4 DISCUSSION

In our study, we investigated the use of the KAN models to generate multiple CEST fitting parameters from Z-spectra. Traditionally, these parameters are extrapolated through MPLF or MLP models. Our results showed that KANs outperformed MLPs in terms of generating more accurate CEST fitting parameters from the same human CEST dataset. Specifically, compared to MLPs, KANs demonstrated lower training and validation losses, higher Pearson coefficients with MPLF, and better accuracy in reconstructed CEST maps and Z-spectra (Figure 2-5). In comparison to the time-consuming MPLF, which took about 5 minutes (for single slice) to process one-slice brain CEST data, both network-based methods could quickly produce results in just a few seconds after training.

KAN offers an innovative approach to network operations by optimizing activation placement and incorporating modular non-linearity[52]. The activation functions in KAN are moved to the edges rather than the neuron core, which potentially alters learning dynamics and enhances interpretability. The applied non-linearity before summing inputs allows for differentiated treatment of features and potentially more precise control over input influence on outputs. The innovative architecture of KAN allows for more effective handling of complex tasks, as seen in CEST MRI data analysis where KAN outperforms MLP. KAN leverages a unique architecture that combines the accuracy of splines with the structural learning capabilities of neural networks. Splines are highly accurate in modeling low-dimensional functions, while neural networks excel in learning compositional structures. This integration facilitates the capture of local details and global patterns in CEST data, where multiple CEST peaks interfere with each other, ultimately enhancing modeling accuracy. However, due to the increased complexity of the network architecture, the training process of KAN was much slower than that of MLP (as seen in Figure 6). Moreover, the training time of KAN showed an obvious dependency on the number of layers. This problem needs to be solved in future studies.

Despite the challenges associated with its training time, KAN offers a highly promising approach for tackling the complex data analysis of CEST MRI. Currently, there is no standardized CEST MRI data analysis method, and while MPLF is a commonly used method, it cannot generate real ground truth for CEST MRI[38,44]. In this study, we used MPLF as an example to demonstrate the feasibility of using KAN for CEST MRI data analysis. We anticipate that KAN can also be applied in other CEST MRI techniques that use MLP in data analysis[39-42]. It also has the potential to be applied in CEST magnetic resonance fingerprinting (MRF) for extracting quantitative CEST fitting parameters, including concentration and exchange rate, from specially designed CEST-MRF spectra[45-51]. Furthermore, it is expected that KAN will have extensive applicability in analyzing data from other MRI techniques that utilize MLP in post-processing.[57-62]

## 5    CONCLUSION

This study demonstrated for the first time the feasibility of using KAN for CEST MRI data analysis in the human brain, with the KAN model surpassing the conventional MLP approach in the accuracy of predicting the multiple CEST fitting parameters. Moreover, similar to other network-based methods, KAN could generate CEST contrast maps at a much faster speed than the conventional MPLF method. These findings suggest that CEST-KAN has the potential to be a robust and reliable post-processing tool for CEST MRI in clinical settings, potentially overcoming the limitations of the current MPLF or MLP-based method.

## ACKNOWLEDGEMENT

Authors would like to thank the funding support from The University of Hong Kong: 109000487, 204610401 and 204610519.

## DATA AVAILABILITY STATEMENT

The multi-pool Lorentzian fitting (MPLF) code for CEST MRI analysis is publicly available at: https://github.com/JianpanHuang/CEST-MPLF. The multi-layer perceptron (MLP) and Kolmogorov-Arnold Network (KAN) codes for CEST MRI analysis are publicly available at: https://github.com/JianpanHuang/CEST-KAN.

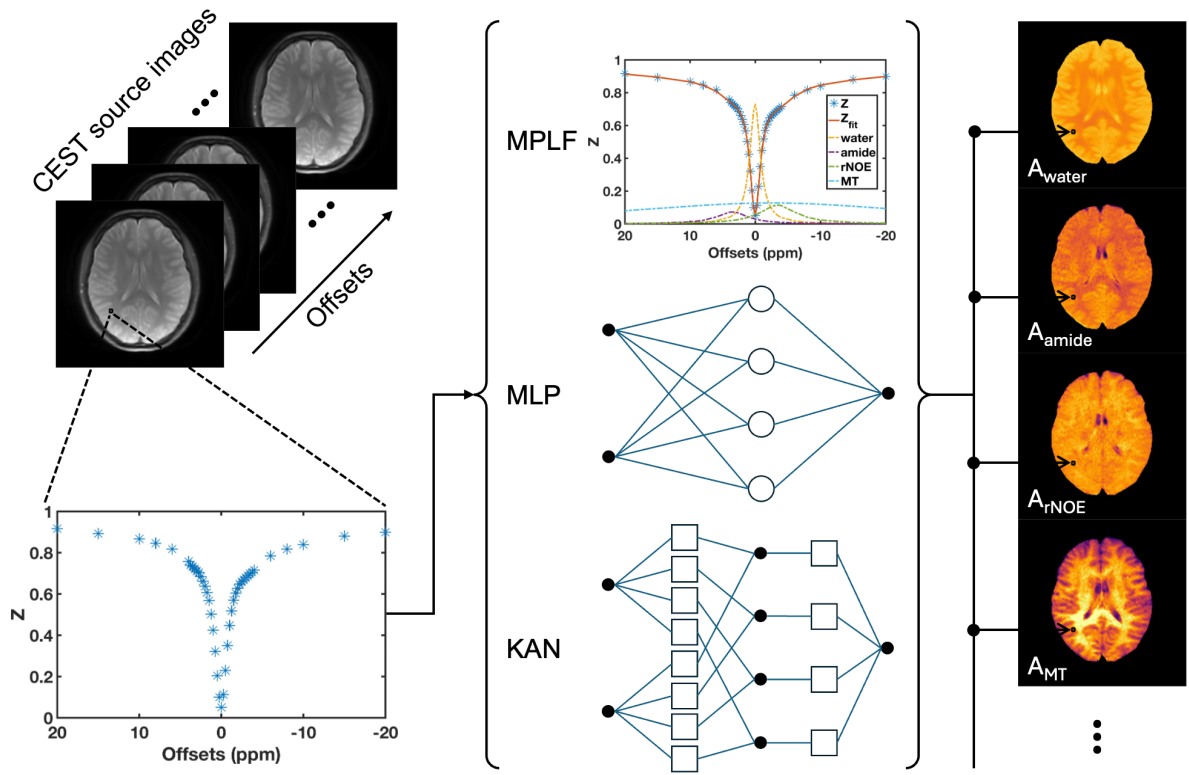

**Figure 1.** Data analysis for CEST MRI using multipool Lorentzian fitting (MPLF), multiple layer perception (MLP) and Kolmogorov-Arnold Network (KAN) methods. The input is the Z-spectrum, and the output is multiple CEST fitting parameters.

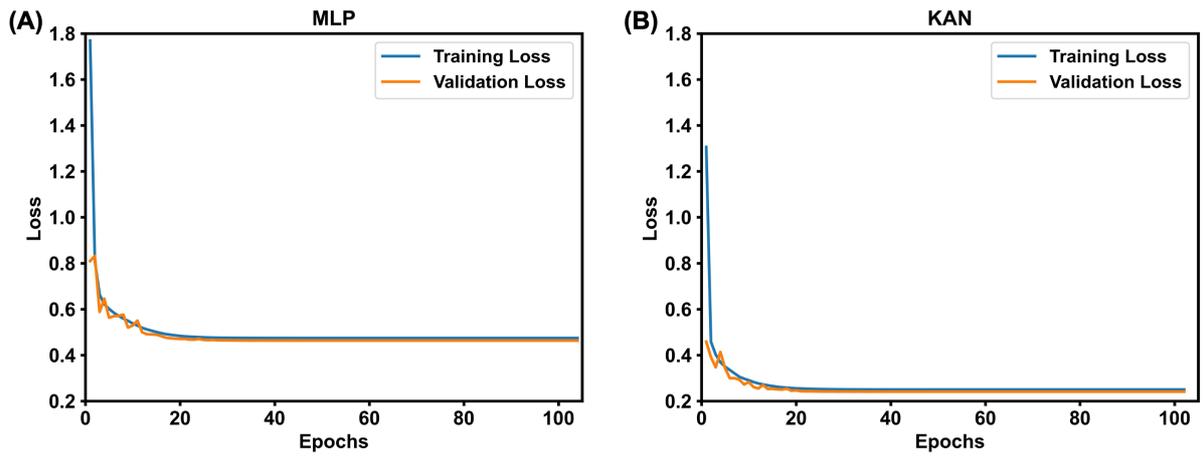

**Figure 2.** Training and validation loss curves of (A) MLP and (B) KAN on the same CEST dataset and with the same parameter settings.

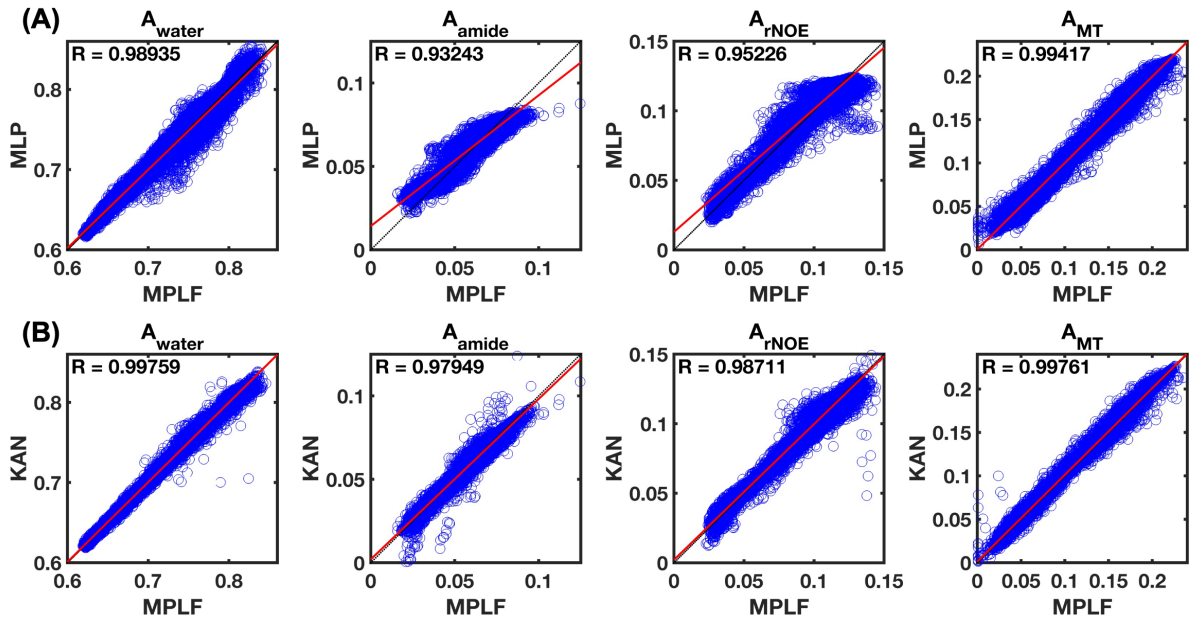

**Figure 3.** Application of the trained MLP and KAN to testing CEST dataset acquired from 2 human subjects. Correlation plots of prediction results of (A) MLP and (B) KAN versus MPLF fitting results. R values in plots denote the Pearson correlation coefficient between prediction and ground truth. The columns represent the corresponding water, amide, rNOE, and MT plots.

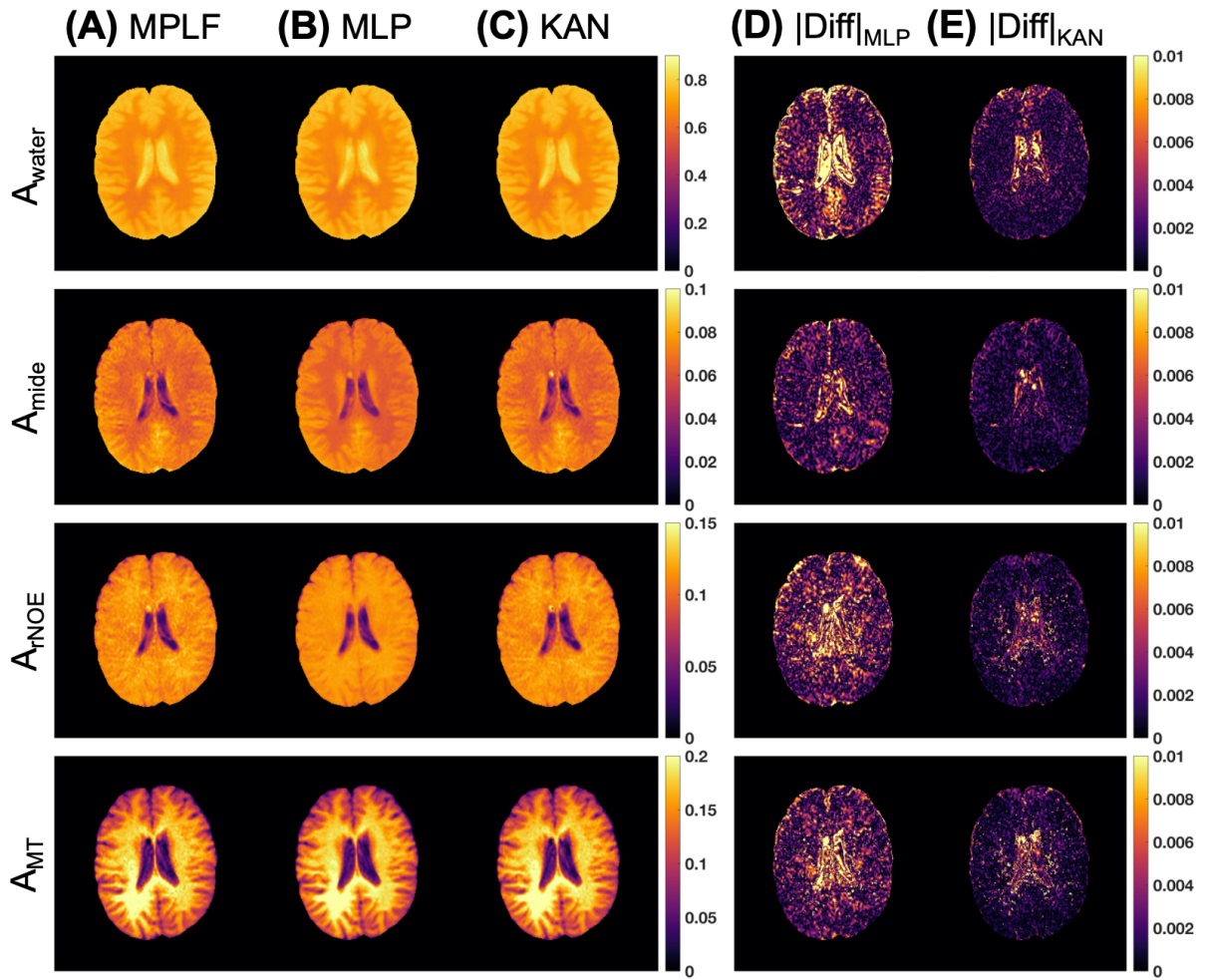

**Figure 4.** Water and CEST maps of MLP and KAN applied to the testing data of subject #1 that was not included in training, compared to MPLF fitting results. (A) Water and CEST maps obtained by MPLF method. (B) Water and CEST maps obtained by MLP method. (C) Water and CEST maps obtained by KAN method. (D) & (E) Absolute difference (|Diff|) between the MPLF fitting outcomes and the predictions of the MLP and KAN models, respectively. The rows represent the corresponding water, amide, rNOE, and MT maps.

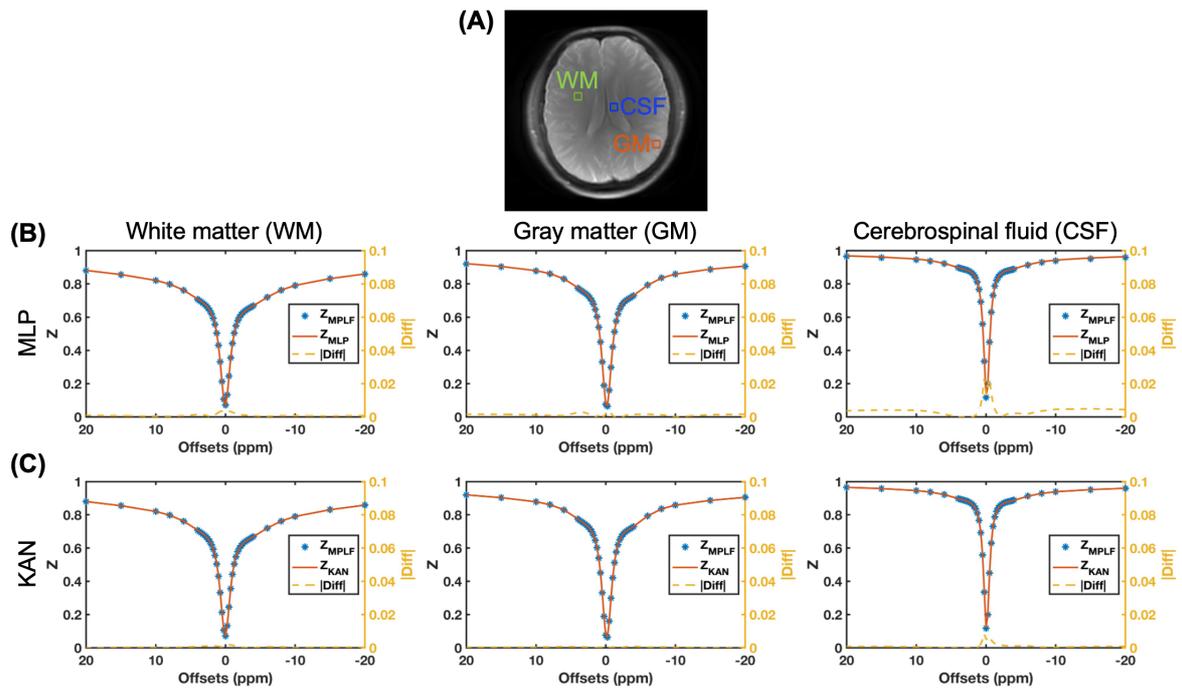

**Figure 5.** Comparison of the Z-spectra in the testing data of subject #1, reconstructed using the CEST fitting parameters extrapolated by MPLF, MLP and KAN. (A) Three ROIs used for generating the Z-spectra of white matter (green), gray matter (orange) and cerebrospinal fluid (blue). (B) Comparison of MLP with MPLF. (C) Comparison of KAN with MPLF.

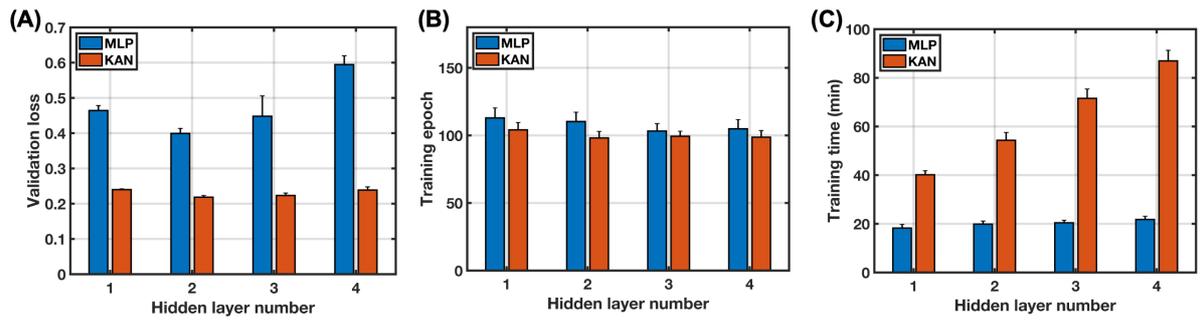

**Figure 6.** Average training results of MLPs and KANs with a fixed layer size of 100 but varing numbers of layers. The average results of 10 trials for each network are depicted. (A) Validation loss. (B) Training epoch. (C) Training time.

**Table 1.** Initial values and bound conditions for MPLF with pool number set to four

| Pool | Parameter | Initial value | Lower bound | Upper bound |
|---|---|---|---|---|
| water | $\Delta\delta$ (ppm) | 0 | -2.0 | 2.0 |
|  | $A_{water}$ | 0.9 | 0.4 | 1.0 |
|  | $\Gamma_{water}$ (ppm) | 2.3 | 1.0 | 6.0 |
| amide | $A_{amide}$ | 0.01 | 0 | 0.25 |
|  | $\Gamma_{amide}$ (ppm) | 2.0 | 1.0 | 12.5 |
| rNOE | $A_{rNOE}$ | 0.1 | 0 | 0.25 |
|  | $\Gamma_{rNOE}$ (ppm) | 4.0 | 2.0 | 15.0 |
| MT | $A_{MT}$ | 0.1 | 0.001 | 0.35 |
|  | $\Gamma_{MT}$ (ppm) | 60 | 30 | 150 |

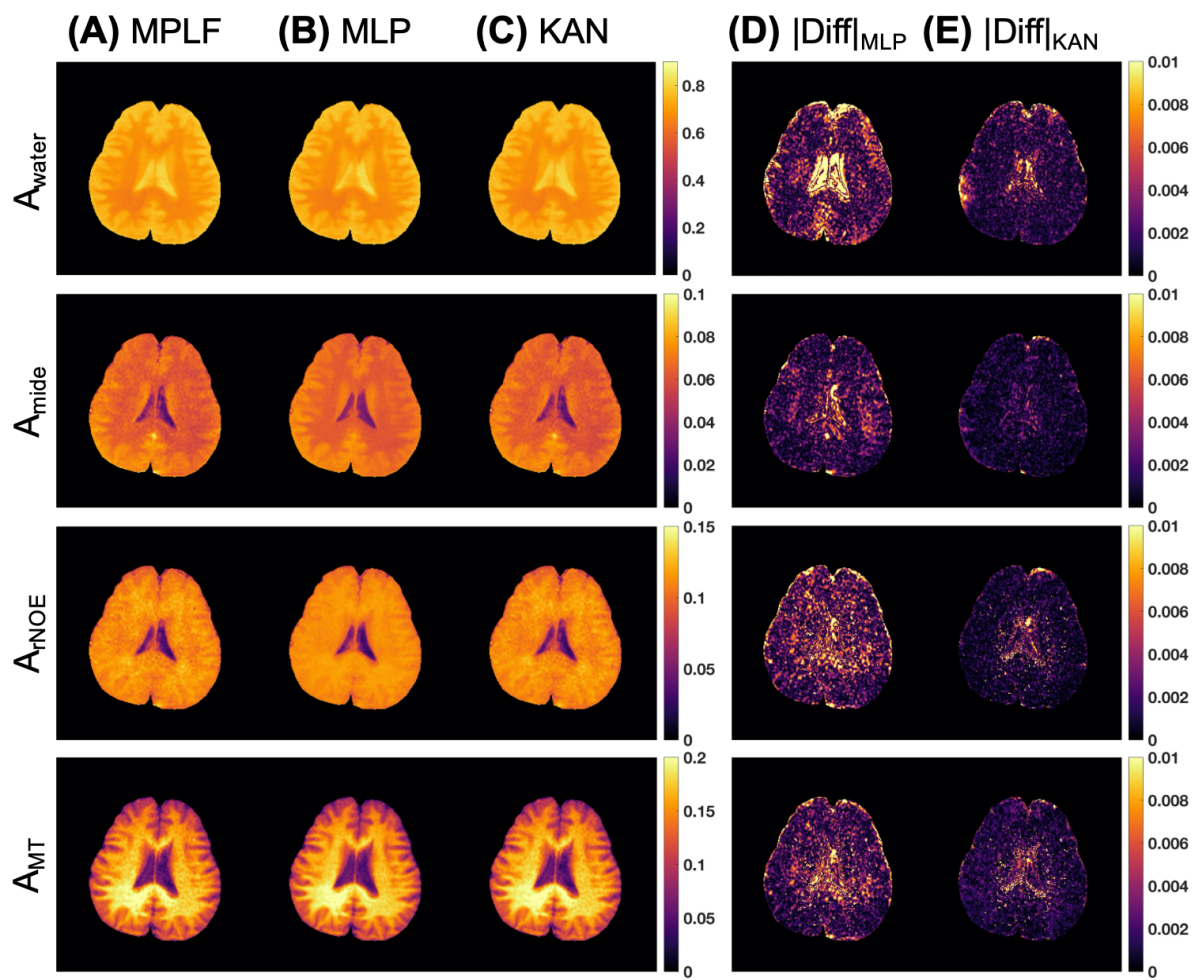

**Figure S1.** Water and CEST maps of MLP and KAN applied to the testing data of subject #2 that was not included in training, compared to MPLF fitting results. (A) Water and CEST maps obtained by MPLF method. (B) Water and CEST maps obtained by MLP method. (C) Water and CEST maps obtained by KAN method. (D) & (E) Absolute difference (|Diff|) between the MPLF fitting outcomes and the predictions of the MLP and KAN models, respectively. The rows represent the corresponding water, amide, rNOE, and MT maps.

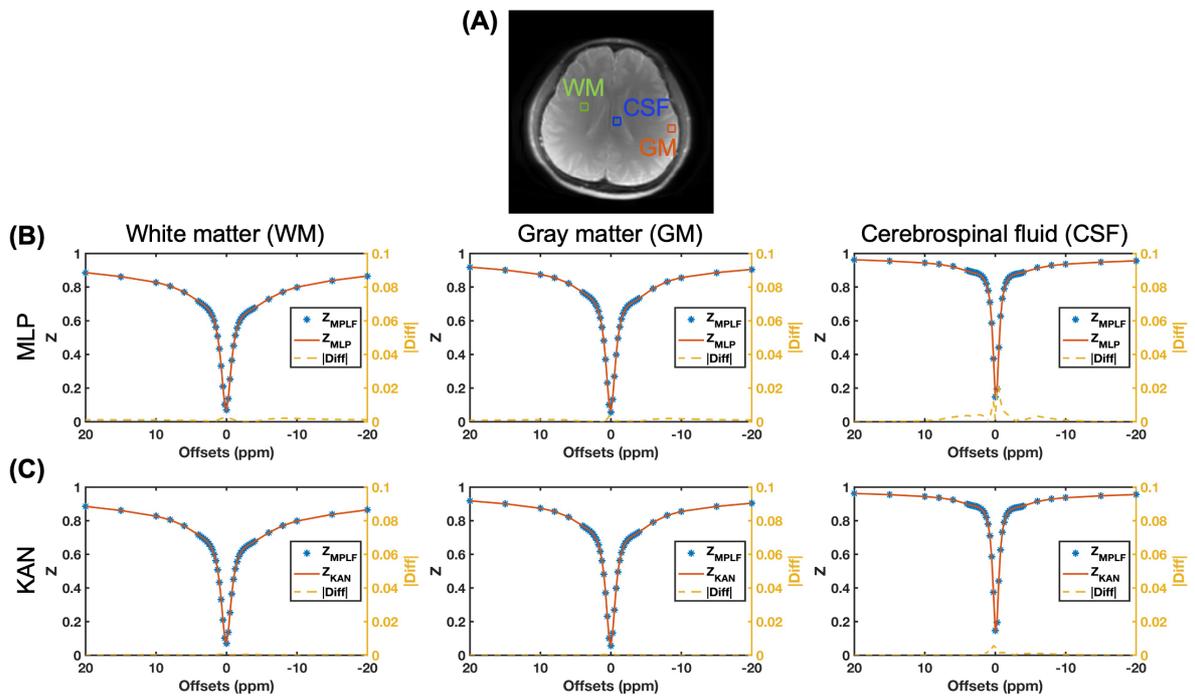

**Figure S2.** Comparison of the Z-spectra in the testing data of subject #2, reconstructed using the CEST fitting parameters extrapolated by MPLF, MLP and KAN. (A) Three ROIs used for generating the Z-spectra of white matter (green), gray matter (orange) and cerebrospinal fluid (blue). (B) Comparison of MLP with MPLF. (C) Comparison of KAN with MPLF.